# Does fractal universe favour warm inflation: Observational support?

Akash Bose, Subenoy Chakraborty *

*Department of Mathematics, Jadavpur University, Kolkata-700032, West Bengal, India*

**Abstract**

The present work examines the possibility of warm inflationary paradigm in the modified gravity theory with fractal geometry. By choosing the normal fluid as radiation fluid and the effective fluid (with the extra term in modified field equation) as the inflaton field both strong and weak dissipative regimes have been studied using slow roll approximation with quasi-stable criteria for radiation. Finally, using the Planck data set, the present model has been analyzed for various choices of the fractal function and the dissipation parameter.

## 1. Introduction

In cosmology, due to tremendous technological development there are series of precision observed data dealing with Cosmic Microwave Background [1,2], large scale structures [3,4], Barionic Acoustic Oscillation data [5,6] both strong and weak lensing [7–9], galaxy cluster number counts [10,11] and gravitational waves detections [12–14]. Based on these observed data one have the idea about the early histories of the universe as well as evolution and nature of the universe [15,16]. Inflationary paradigm is the best predictive description of the universe at the early eras just after big bang. This scenario not only solves the problems of the standard big bang cosmology but also it explains the origin of the CMB anisotropies and the large scale structure of the universe [17,18]. On the other hand, recent CMB data discards several models of inflation and puts severe constraints on many other models [19] with single scalar field (inflaton) model is the best option.

* Corresponding author.
  *E-mail addresses:* bose.akash13@gmail.com (A. Bose), schakraborty.math@gmail.com (S. Chakraborty).

At present, there are two distinct inflationary scenarios depending on the dynamics of the inflaton field and they are termed as cold and warm inflationary models. In cold inflationary (CI) mechanism the inflaton field interacts with other field degrees of freedom very weakly so that it cannot prevent the dilution of any pre-existing or newly formed radiation and as a result there is a period of supercooling in this scenario. Also in this model density perturbations are originated from the quantum fluctuations of the inflaton field [20]. On the other hand, in warm inflationary (WI) model [21–23] the interaction between the inflaton and other fields is very much dominated to produce a quasi-stationary thermalized radiation bath during inflation. So the thermal fluctuations in the radiation bath is the primary source of density fluctuations and it is transported to the inflaton field as adiabatic curvature perturbations [24–28]. Due to this non-trivial dissipative dynamics (and also stochastic effects) the observed inflationary parameters namely tensor to scalar ratio ($r$), the scalar spectral index ($n_s$) and the non-Gaussianity parameters ($f_{NL}$) [29–34] differ significantly from their values in CI. Further, in WI scenario it is possible to have a strong coupling between the inflaton field and other fields to have sufficient amount of radiation production, preserving the required flatness of the potential. As a result, the supercooling of the universe (observed in CI) is compensated by the radiation production and the universe makes a smooth transition from the accelerated era of expansion (inflationary epoch) to radiation dominated phase in WI without encountering any (pre) reheating era. Moreover, according to swampland conjectures [35–37], it is not possible to have de-Sitter vacua in string theory and also set very stringent constraints on inflation model-building leading to impossibility for CI while dissipation mechanism dominated WI can accommodate these conjectures.

The paper is organized as follows: A brief review of warm inflation has been done in section 2. The basics of fractal gravity theory has been briefly discussed in section 3. Section 4 deals with a detailed study of warm inflation in fractal geometry. In this section different choices for fractal function have been made to explain the inflationary paradigm. Also for that choice of fractal function, both weak and strong dissipative regime have been investigated with two different choices of dissipation coefficient. Several physical parameters, for example slow roll parameters, tensor to scalar ratio ($r$), scalar spectral index ($n_s$) have been formulated as a function of model parameters. Then this theoretical model has been compared with the $r - n_s$ diagram of the Planck 2018 data to find out an admissible range of the model parameters. The paper ends with summary and conclusion in section 5.

## 2. Basic equations in warm inflationary scenario

In the background of flat FLRW space-time geometry, the Friedmann equations for warm inflation can be written as

$$3H^2 = \rho_\phi + \rho_r, \qquad 2\dot{H} + 3H^2 = -p_\phi - p_r \tag{1}$$

where ($\rho_r$, $p_r$) are the energy density and thermodynamic pressure of the radiation component and those for scalar field are $\rho_\phi$ and $p_\phi$ respectively.

Due to interaction between these two components, scalar field decays and consequently energy is transferred from scalar field to radiation fluid, and this process is described by the following energy conservation equations,

$$\dot{\rho}_r + 3H(\rho_r + p_r) = \Gamma\dot{\phi}^2, \quad (2)$$

$$\dot{\rho}_\phi + 3H(\rho_\phi + p_\phi) = -\Gamma\dot{\phi}^2, \quad (3)$$

where $\Gamma$ is the dissipation coefficient and it is considered as positive according as the second law of thermodynamics. By using the explicit expressions for the energy density and pressure of the scalar field (i.e. $\rho_\phi = \frac{1}{2}\dot{\phi}^2 + V(\phi)$, $p_\phi = \frac{1}{2}\dot{\phi}^2 - V(\phi)$), equation (3) can be rewritten as the generalized Klein-Gordon equation for the scalar field,

$$\ddot{\phi} + 3H(1+Q)\dot{\phi} + \frac{\partial V(\phi)}{\partial \phi} = 0 \quad (4)$$

where the parameter $Q = \frac{\Gamma}{3H}$ is the ratio of the radiation production to expansion rate. By assuming the process to be quasi-de Sitter, the scalar field energy density dominates over the radiation energy density, $\rho_\phi \gg \rho_r$, and the kinetic term of the scalar field is negligible compared to the potential, i.e. $\rho_\phi \simeq V(\phi)$. Also it is assumed that the production of the radiation component is quasi-stable during inflation, so $\dot{\rho}_r \ll H\rho_r$ and $\dot{\rho}_r \ll \Gamma\dot{\phi}^2$. Then equations (1)-(4) can be approximated as

$$3H^2 \simeq V(\phi) \quad (5)$$

$$\dot{\phi} \simeq -\frac{V'(\phi)}{3H(1+Q)} \quad (6)$$

$$\rho_r \simeq \frac{\Gamma\dot{\phi}^2}{4H} = C_\gamma T^4 \quad (7)$$

where $T$ is the temperature of the radiation bath, $C_\gamma = \frac{\pi^2 g_*}{30}$ is the Stefan-Boltzmann constant, and $g_*$ is the number of degrees of freedom of the radiation fluid. Here 'overdot' and 'prime' denotes the differentiation with respect to cosmic time $t$ and scalar field $\phi$ respectively.

Now during warm inflationary scenario, one can identify two regimes: namely the weak dissipative regime where $Q \ll 1$ (equivalently $\Gamma \ll 3H$) and the strong dissipative regime in which $Q \gg 1$ (i.e., $\Gamma \gg 3H$). Moreover, one can consider the parameter $\Gamma$ to be a constant (i.e., $\Gamma_0$) or a function of the potential (i.e., $\Gamma = \Gamma(\phi)$) or a function of the temperature of thermal bath $T$ (i.e., $\Gamma = \Gamma(T)$) or function of both (i.e., $\Gamma = \Gamma(\phi, T)$). If one consider $\Gamma$ as a function of both $\phi$ and $T$, then using the result of quantum field theory in curved space, $\Gamma$ can be chosen as $\Gamma = \Gamma_0 \frac{T^m}{\phi^{m-1}}$.

Using the above equations, one can express the first slow roll parameter as

$$\epsilon_1 = \frac{1}{2(1+Q)}\left(\frac{V'}{V}\right)^2 = -\frac{\dot{H}}{H^2} \quad (8)$$

The other two slow roll parameters can be expressed as

$$\eta = \frac{1}{(1+Q)}\frac{V''}{V}, \quad \beta = \frac{1}{(1+Q)}\frac{V'\Gamma'}{V\Gamma} \quad (9)$$

The amount of cosmic expansion during the inflationary epoch is measured by the number of e-folds, $N$, given by

$$N = \int_{t_\star}^{t_e} H dt = \int_{\phi_\star}^{\phi_e} \frac{H}{\dot{\phi}} d\phi = -\int_{\phi_\star}^{\phi_e} (1+Q) \frac{V}{V'} d\phi \qquad (10)$$

where the subscripts '$e$' and '$\star$' denote the quantity at the end of the inflation and at the horizon crossing time respectively.

On the other hand, due to the presence of radiation fluid in warm inflation dynamics, the source of density fluctuations correspond to thermal fluctuations. Thermal fluctuations depend on the fluid temperature $T$ while quantum fluctuations are dependent on the Hubble parameter $H$. In the warm inflationary scenario, the fluid temperature is larger than the Hubble parameter, i.e. $T > H$; stating that the thermal fluctuations are dominant over the quantum fluctuations, and become the origin of the Universe's LSS. The amplitude of scalar perturbation during warm inflation, along with the slow roll approximation becomes

$$P_s = \frac{H^3 T}{\dot{\phi}^2} (1+Q)^{\frac{1}{2}} = \frac{HT}{2\epsilon_1} (1+Q)^{\frac{3}{2}} \qquad (11)$$

The scalar spectral index $n_s$ [38] is defined as $n_s - 1 = \dfrac{d \ln P_s}{d \ln k}$ and can be written in terms of slow roll parameters as

$$n_s - 1 = -\frac{(9Q+17)}{4(1+Q)} \epsilon_1 - \frac{(9Q+1)}{4(1+Q)} \beta + \frac{3}{2} \eta \qquad (12)$$

According to the latest observational data, the amplitude of the scalar perturbations at horizon crossing is given by $P_s = 2.17 \times 10^{-9}$ [39,40]. The amplitude of tensor perturbation is given by $P_t = 8H^2$. The tensor to scalar ratio $r$ is given by

$$r = \frac{P_t}{P_s} = \frac{16\epsilon_1}{(1+Q)^{\frac{3}{2}}} \frac{H}{T} \qquad (13)$$

## 3. Brief review of fractal gravity

The total action of the Einstein gravity in fractal space-time is given by

$$S = S_g + S_m, \qquad (14)$$

where the gravitational part of the action is given by

$$S_g = \frac{1}{16\pi G} \int d^4x \, v(x) \sqrt{-g} \left( R - \omega \partial_\mu v \partial^\mu v \right), \qquad (15)$$

and $S_m$, the action of the matter part minimally coupled to gravity [41–43], is given by

$$S_m = \int d^4x \, v(x) \sqrt{-g} \mathcal{L}_m. \qquad (16)$$

Here $g$ is the determinant of the metric $g_{\mu\nu}$, $R$ is the Ricci scalar, $v$ is the fractal function and $\omega$ is the fractal parameter. The standard measure $d^4x$ is replaced by a Lebesgue-Stieltjes measure $d\mathfrak{g}(x)$. The scaling dimension of $\mathfrak{g}$ is $-4\alpha$, where the parameter $\alpha$ ($0 < \alpha < 1$) corresponds to the fraction of states preserved at a given time during the evolution of the system. Further, the measure $v$ is not a scalar field (nor a dynamical object), but its profile is fixed a priori by the underlying geometry.

Now varying the above action (14), with respect to homogeneous, isotropic and flat FLRW metric $g_{\mu\nu}$, the Friedmann equations in a fractal Universe can be obtained as (for convenience $8\pi G = 1$ is chosen)

$$3H^2 = \rho - 3H\frac{\dot{v}}{v} + \frac{\omega}{2}\dot{v}^2 \tag{17}$$

$$2\dot{H} + 3H^2 = -p - 2H\frac{\dot{v}}{v} - \frac{\omega}{2}\dot{v}^2 - \frac{\ddot{v}}{v} \tag{18}$$

where $H = \frac{\dot{a}}{a}$ is the Hubble parameter, $a$ is the scale factor, $\rho$ and $p$ are the energy density and pressure of the barotropic fluid component, respectively, with the barotropic equation of state $p = \omega\rho$. One may note that if the fractal function $v$ is chosen as constant, then the standard Friedmann equations are recovered. The continuity equation in a fractal universe takes the form [44–47]

$$\dot{\rho} + \left(3H + \frac{\dot{v}}{v}\right)(p + \rho) = 0 \tag{19}$$

The above modified Einstein field equations can be written as Einstein field equations with interacting two fluids as [44,46]

$$3H^2 = \rho + \rho_f$$
$$2\dot{H} + 3H^2 = -p - p_f$$

with conservation equations

$$\dot{\rho} + 3H(\rho + p) = Q = -\frac{\dot{v}}{v}(\rho + p),$$
$$\dot{\rho}_f + 3H(\rho_f + p_f) = -Q$$

and $\rho_f = \frac{\omega}{2}\dot{v}^2 - 3H\frac{\dot{v}}{v}$, $p_f = 2H\frac{\dot{v}}{v} + \frac{\omega}{2}\dot{v}^2 + \frac{\ddot{v}}{v}$

## 4. WI in fractal gravity

In the context of present WI scenario, $(\rho_f, p_f)$ can be considered as the energy density and thermodynamic pressure for the inflaton field (i.e. $\rho_f = \rho_\phi = \frac{1}{2}\dot{\phi}^2 + V(\phi)$, $p_f = p_\phi = \frac{1}{2}\dot{\phi}^2 - V(\phi)$) while the usual field is chosen as radiation field i.e. $\rho = \rho_r$, $p = p_r = \frac{1}{3}\rho_r$. So the evolution equations of two fluid can be separately written as

$$\dot{\rho}_r + 4H\rho_r = \Gamma\dot{\phi}^2 = -\frac{4}{3}\frac{\dot{v}}{v}\rho_r, \tag{20}$$

$$\dot{\rho}_\phi + 3H(\rho_\phi + p_\phi) = -\Gamma\dot{\phi}^2 = \frac{4}{3}\frac{\dot{v}}{v}\rho_r. \tag{21}$$

It is to be noted that the nonminimal interaction term $\Gamma$ in equations (20) and (21) is not phenomenological; rather, it is a consequence of the continuity equation (19). In particular, in the present model, $\Gamma$ depends on the time variation of the logarithm of $v$.

Assuming the quasi stable production of the radiation component (i.e. $\dot{\rho}_r \ll H\rho_r$ and $\dot{\rho}_r \ll \Gamma\dot{\phi}^2$), one can obtain the Hubble parameter by using equation (20) as

$$H = -\frac{1}{3}\frac{\dot{v}}{v} \tag{22}$$

and hence the potential can also be written in terms of fractal function as

$$V(\phi) = \frac{1}{3}\left(\frac{\dot{v}}{v}\right)^2 \tag{23}$$

Now, we choose various choices of the fractal function.

4.1. Model I: $v = v_0 t^{-\delta}$, $\delta = 4(1-\alpha)$

This is the most common choice for the fractal function in literature [41,44]. Moreover, it neither represents any multifractal geometry nor it recovers the standard measure of GR. However, the profile $v = 1 + t^n$ has the best theoretical motivation, but we have not considered it due to complexity of calculation [45]. From equation (22) one can obtain the Hubble parameter, and hence the scale factor, and the potential function (from equation (23)) can be expressed in terms of cosmic time $t$ as

$$H = \frac{\delta}{3}\frac{1}{t}, \qquad a = a_0 t^{\frac{\delta}{3}}, \qquad V(\phi) = \frac{\delta^2}{3}\frac{1}{t^2} \tag{24}$$

Thus the 1st slow roll parameter turns out to be constant as

$$\epsilon_1 = \frac{3}{\delta}$$

Hence one has power law inflation for the choice of the fractal function and there is no mechanism to halt the inflation, it continues for ever.

4.2. Model II: $v = v_0 e^{-\delta t}$, $\delta = 4(1-\alpha)$

In literature some authors choose this type of fractal function. As in the previous model, it neither represents any multifractal geometry nor it recovers the anomalous scaling of a fractal geometry. In this case equation (22) leads to

$$H = \frac{\delta}{3}, \qquad a = a_0 \exp\left(\frac{\delta}{3}t\right), \qquad V(\phi) = \frac{\delta^2}{3} \tag{25}$$

and consequently one can see that the first slow roll parameter vanishes.

So for this choice of the fractal function the universe corresponds to de-Sitter model.

4.3. Model III: $v = v_0 e^{-\delta t^n}$, $\delta = 4(1-\alpha)$

This model can be regarded as a nonperturbative extension of the measure $v = 1 + t^n$ [42] (Note that the constant 1 is necessary to recover GR). Moreover, expanding the measure of the present model, one gets $v \approx \text{constant} + t^n$ which has the correct multiscaling. However, this would happen only at early times (not at late times) or when $\alpha$ is very close to 1 (which is not theoretically desirable). To obtain non-constant slow roll parameters for inflationary paradigm, one can choose phenomenologically the fractal function in the above form by generalizing the choice for model II. Similarly from equation (22) and (23) one has

$$H = \frac{\delta n}{3} t^{n-1}, \qquad a = a_0 \exp\left(\frac{\delta}{3} t^n\right), \qquad V(\phi) = \frac{\delta^2 n^2}{3} t^{2n-2} \qquad (26)$$

The slow roll parameter can also be obtained as

$$\epsilon_1 = \frac{3(1-n)}{\delta n} t^{-n} \qquad (27)$$

It is to be noted that, the value of $\alpha$ should be close to 0 i.e., $\delta$ is close to 1; so that the first slow roll parameter $\epsilon_1$ (given in (27)) becomes realistic (i.e., $\epsilon_1 \ll 1$). However, if $\alpha$ is close to 1, i.e., $\delta$ is close to 0, then $\epsilon_1$ can no longer be very small and hence slow roll approximation cannot be valid. Hence, it is not physically justified. Also very small value of $\alpha$, implies the fractal dimension should be very small.

### 4.3.1. Weak dissipative regime

In the weak dissipative regime, i.e. $Q \ll 1$, one can write the evolution equation of the scalar field as

$$\dot{\phi}^2 = -\frac{\dot{V}}{3H} = \frac{2\delta n(1-n)}{3} t^{n-2} \qquad (28)$$

which gives the scalar field explicitly as

$$\phi = \sqrt{\frac{8\delta(1-n)}{3n}} t^{\frac{n}{2}} = \phi_0 t^{\frac{n}{2}} \qquad (29)$$

and the potential can be obtained as

$$V(\phi) = \frac{\delta^2 n^2}{3} \left(\frac{8\delta(1-n)}{3n}\right)^{\frac{2(1-n)}{n}} \phi^{\frac{4(n-1)}{n}} = V_0 \phi^{\frac{4(n-1)}{n}} \qquad (30)$$

Hubble parameter, scale factor and the fractal function can be expressed in terms of scalar field as

$$H(\phi) = H_0 \phi^{\frac{2(n-1)}{n}} \text{ with } H_0 = \frac{\delta n}{3} \left(\frac{8\delta(1-n)}{3n}\right)^{\frac{(1-n)}{n}} \qquad (31)$$

$$a(\phi) = a_0 \exp\left(\frac{n}{8(1-n)} \phi^2\right) \qquad (32)$$

One may note that $V_0 = 3H_0^2$. The slow roll parameters can be obtained as

$$\epsilon_1(\phi) = \frac{8(n-1)^2}{n^2 \phi^2} \qquad (33)$$

$$\eta(\phi) = \frac{4(n-1)(3n-4)}{n^2 \phi^2} \qquad (34)$$

At the end of inflation, $\epsilon_1(\phi_e) = 1$. Therefore, $\phi_e^2 = \frac{8(n-1)^2}{n^2}$. The number of e-folds is given by

$$N = -\int_{\phi_\star}^{\phi_e} \frac{V}{V'} d\phi = \frac{n}{8(1-n)} \left[\frac{8(n-1)^2}{n^2} - \phi_\star^2\right] \qquad (35)$$

which consequently gives the scalar field at the time of horizon crossing as

$$\phi_\star^2 = \frac{8(n-1)^2}{n^2}\left[1 + \frac{n}{n-1}N\right] \tag{36}$$

The slow roll parameters at the horizon crossing are

$$\epsilon_{1\star} = \left(1 + \frac{n}{n-1}N\right)^{-1} \tag{37}$$

$$\eta_\star = \frac{(3n-4)}{2(n-1)}\left(1 + \frac{n}{n-1}N\right)^{-1} \tag{38}$$

**Case 1.** $\Gamma = \Gamma_0 \phi^m$

The temperature of the radiation fluid can be obtained as

$$T = \left(\frac{\Gamma_0(1-n)}{2C_\gamma}\right)^{\frac{1}{4}} \phi_0^{\frac{1}{2n}} \phi^{\frac{1}{4}(m-\frac{2}{n})} \tag{39}$$

The other slow roll parameter can be obtained as

$$\beta(\phi) = \frac{4m(n-1)}{n\phi^2} \tag{40}$$

and at the horizon crossing time it is given by

$$\beta_\star = \frac{mn}{2(n-1)}\left(1 + \frac{n}{n-1}N\right)^{-1} \tag{41}$$

Substituting the above values of slow roll parameters, the scalar spectral index $n_s$ can be written as a function of number of e-folds as

$$n_{s_\star}(n,m,N) = 1 + \frac{n}{4(1-n)}\left(8 + \frac{m}{2} - \frac{5}{n}\right)\left(1 + \frac{n}{n-1}N\right)^{-1} \tag{42}$$

The tensor to scalar ratio at the same time can be written as

$$r_\star(n,m,N) = \frac{16n\delta}{3}\left(\frac{2C_\gamma}{1-n}\right)^{\frac{1}{4}} \frac{1}{\Gamma_0^{\frac{1}{4}}(n,m,N)} \left(\frac{3n}{8\delta(1-n)}\right)^{1-\frac{3}{4n}} \left(\frac{8(n-1)^2}{n^2}\right)^{1-\frac{m}{8}-\frac{3}{4n}}$$
$$\times \left(1 + \frac{n}{n-1}N\right)^{-\frac{m}{8}-\frac{3}{4n}} \tag{43}$$

Using the definition of $P_s$, one can find out $\Gamma_0(n,m,N)$ and the tensor to scalar ratio can be simplified to

$$r_\star(n,m,N) = \frac{8(1-n)^2}{P_s}\left(\frac{\delta n}{3(1-n)}\right)^{\frac{2}{n}}\left(1 + \frac{n}{n-1}N\right)^{2-\frac{2}{n}} \tag{44}$$

Using the $r - n_s$ diagram of Planck-2018, one could plot a $n - m$ diagram as shown in Fig. 1, where the dark blue colour and light blue colour indicate an area of $(n,m)$ in which the point $(r, n_s)$ of the model stand in 68% and 95% CL respectively.

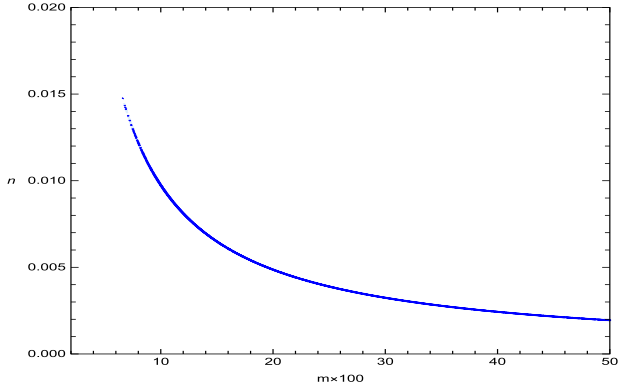

Fig. 1. Numerical values of the $(n, m)$ parameters of the fractal warm inflation model in the weak dissipative regime for which the point $(r - n_s)$ is located in the observational region.

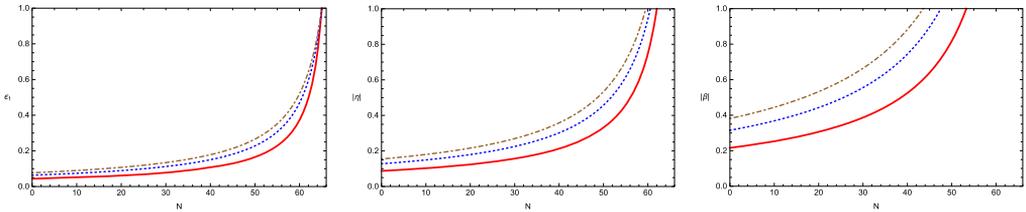

Fig. 2. Evolution of slow roll parameters with respect to the number of e-folds.

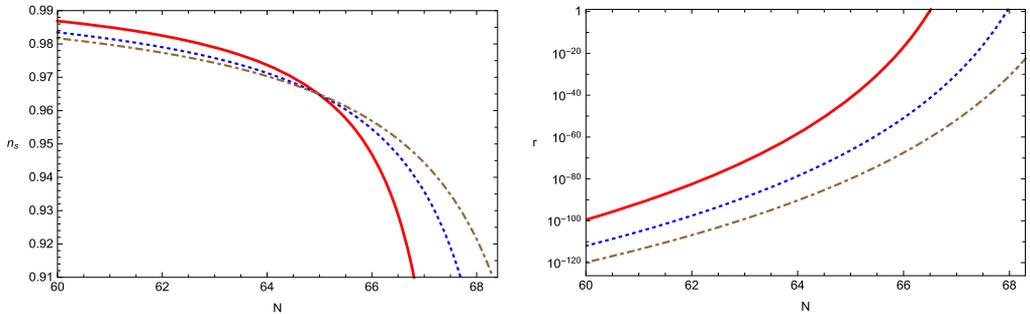

Fig. 3. Variation of spectral index and tensor to scalar ratio in terms of no. of e-fold.

Considering some points of $(n, m)$ from Fig. 1, the evolution of the slow roll parameters with respect to the number of e-folds have been shown in Fig. 2. The figures show that it support the slow roll approximations. The behaviour of the $n_s$ and $r$ have been represented in Fig. 3. The ratio of temperature and Hubble parameter has been depicted during inflation in WDR and have been presented in Fig. 4. The Figs. 2-4 have been drawn considering the data sets for $(n, m)$ as $(.0145, 672.82)$ (solid line), $(.0142, 686.98)$ (dotted line) and $(.014, 696.76)$ (dot-dashed line).

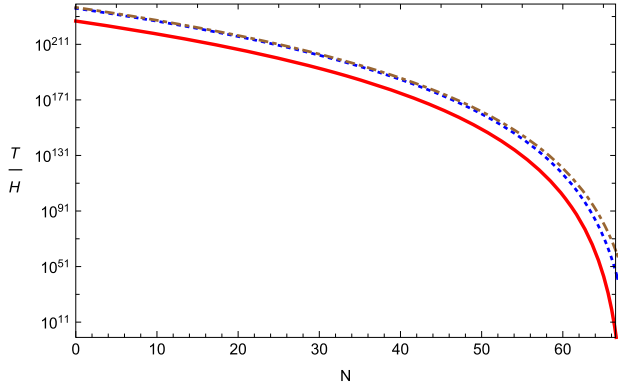

Fig. 4. Ratio of temperature and Hubble parameter in terms of no. of e-fold.

**Case 2:** $\Gamma = \Gamma_0 \dfrac{T^m}{\phi^{m-1}}$

The temperature can be expressed as

$$T^{m-4} = \frac{2C_\gamma}{\Gamma_0(1-n)} \left(\frac{3n}{8\delta(1-n)}\right)^{\frac{1}{n}} \phi^{m-1+\frac{2}{n}} \qquad (45)$$

$\Gamma$ can be explicitly written as

$$\Gamma = \Gamma_0 \left(\frac{2C_\gamma}{\Gamma_0(1-n)}\right)^{\frac{m}{m-4}} \left(\frac{3n}{8\delta(1-n)}\right)^{\frac{m}{n(m-4)}} \phi^{\frac{2(m-2n+2mn)}{n(m-4)}} \qquad (46)$$

The slow roll parameter $\beta$ can be written as

$$\beta(\phi) = \frac{4(n-1)}{n^2} \frac{(m-2n+2mn)}{(m-4)} \frac{1}{\phi^2} \qquad (47)$$

The scalar spectral index can be written as

$$n_{s_\star}(n,m,N) = 1 + \frac{9m + 66n - 18mn - 40}{8(m-4)(n-1)} \left(1 + \frac{n}{n-1}N\right)^{-1} \qquad (48)$$

The tensor to scalar ratio can be written as

$$r_\star(n,m,N) = \frac{16n\delta}{3} \left(\frac{1-n}{2C_\gamma}\right)^{\frac{1}{m-4}} \Gamma_0^{\frac{1}{m-4}}(n,m,N) \left(\frac{3n}{8\delta(1-n)}\right)^{\frac{mn-m-4n+3}{n(m-4)}}$$

$$\left(\frac{8(n-1)^2}{n^2}\right)^{\frac{mn-2m-7n+6}{2n(m-4)}} \left(1 + \frac{n}{n-1}N\right)^{\frac{n-2m-mn+6}{2n(m-4)}} \qquad (49)$$

In Fig. 5, $n - m$ diagram has been plotted using the $r - n_s$ diagram of Planck-2018 data as before, in the figure the dark blue colour indicates an area of $(n, m)$ with the points $(r, n_s)$ of the model in 68% CL while that with light blue colour stands for 95% CL.

One can see that the equation (46) can also be rewritten as

$$\Gamma = \hat{\Gamma}_0 \phi^{\hat{m}}$$

which implies that case 2 can be reduced to case 1 with some modified values of $n$ and $m$. So further detailed study of graphical representation has not been done in this section.

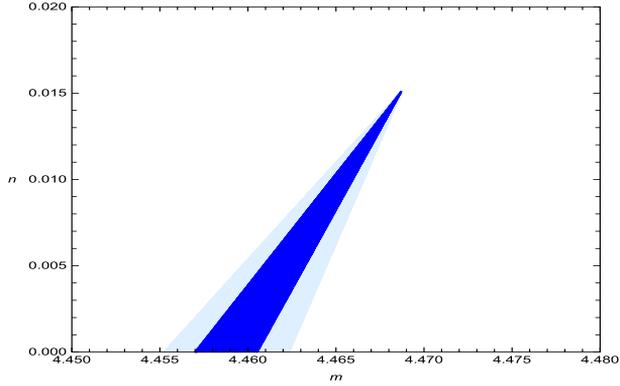

Fig. 5. Observed region indicates the location of $(n, m)$ points in weak dissipative regime for present warm inflationary model considering the $(r, n_s)$ in observational region.

### 4.3.2. Strong dissipative regime

In the strong dissipative regime, the dissipative ratio is much larger than unity ($Q \gg 1$). Using this approximation, one can write the equation (6) as

$$\dot{\phi} = -\frac{\frac{\partial V(\phi)}{\partial \phi}}{\Gamma} \tag{50}$$

**Case 1.** $\Gamma = \Gamma_0 \phi^m$

The evolution equation for the scalar field can explicitly be written as

$$\dot{\phi}^2 = \frac{2\delta^2 n^2 (1-n)}{3\Gamma_0} \frac{t^{2n-3}}{\phi^m} \tag{51}$$

and subsequently on integration it gives

$$\phi^{m+2} = \left(\frac{m+2}{2n-1}\right)^2 \frac{2\delta^2 n^2 (1-n)}{3\Gamma_0} t^{2n-1} = \phi_0 t^{2n-1} \tag{52}$$

Using the above relation, one can write the potential, Hubble parameter, scale factor as

$$V(\phi) = \frac{\delta^2 n^2}{3\phi_0^{\frac{2n-2}{2n-1}}} \phi^{\frac{(m+2)(2n-2)}{(2n-1)}} \tag{53}$$

$$H(\phi) = \frac{\delta n}{3\phi_0^{\frac{n-1}{2n-1}}} \phi^{\frac{(m+2)(n-1)}{(2n-1)}} \tag{54}$$

$$a(\phi) = a_0 \exp\left(\frac{\delta}{3} \frac{\phi^{\frac{n(m+2)}{2n-1}}}{\phi_0^{\frac{n}{2n-1}}}\right) \tag{55}$$

Temperature can be obtained as

$$T = \left[\frac{(1-n)n\delta}{3C_\gamma}\right]^{\frac{1}{4}} \left[\frac{\phi^{m+2}}{\phi_0}\right]^{\frac{n-2}{4(2n-1)}} \tag{56}$$

The slow roll parameters can be written as

$$\epsilon_1 = \frac{3(1-n)}{\delta n} \frac{\phi^{\frac{(m+2)n}{1-2n}}}{\phi_0^{\frac{n}{1-2n}}} \qquad (57)$$

$$\eta = \frac{(2m - 2n - 2mn + 3)}{(m+2)(1-n)} \epsilon_1 \qquad (58)$$

$$\beta = \frac{m(1-2n)}{(m+2)(1-n)} \epsilon_1 \qquad (59)$$

Inflation ends with $\epsilon_1(\phi_e) = 1$ which implies $\phi_e^{\frac{(m+2)n}{(2n-1)}} = \frac{3(1-n)}{\delta n} \phi_0^{\frac{n}{2n-1}}$. Using the above relation for $N$ one can write the scalar field at the horizon crossing explicitly as

$$\phi_\star^{\frac{(m+2)n}{(2n-1)}} = \frac{3(1-n)}{\delta n} \phi_0^{\frac{n}{2n-1}} \left(1 + \frac{n}{n-1} N\right)$$

The slow roll parameters at the horizon crossing can be expressed as

$$\epsilon_{1\star} = \left(1 + \frac{n}{n-1} N\right)^{-1} \qquad (60)$$

$$\eta_\star = \frac{(2mn - 2m + 2n - 3)}{(m+2)(n-1)} \left(1 + \frac{n}{n-1} N\right)^{-1} \qquad (61)$$

$$\beta_\star = \frac{m(2n-1)}{(m+2)(n-1)} \left(1 + \frac{n}{n-1} N\right)^{-1} \qquad (62)$$

Inserting the above parameters, the scalar spectral index at the same time can be obtained as

$$n_s(n, m, N) = 1 + \frac{3}{4} \frac{(2m - 2n - 5mn)}{(m+2)(n-1)} \left(1 + \frac{n}{n-1} N\right)^{-1} \qquad (63)$$

In Fig. 6, at 68% and 95% CL, the points $(r, n_s)$ of the model are indicated by dark blue colour and light blue colour respectively in the $n - m$ plot, considering $(r - n_s)$ diagram of Planck 2018 data.

Choosing some points of $(n, m)$ from Fig. 6, the evolution of the slow roll parameters with respect to the number of e-folds have been plotted in Fig. 7 and it matches with the slow roll approximations. The behaviour of the $n_s$ and $r$ have been represented in Fig. 8. The Figs. 7-8 have been drawn considering the values for $(n, m)$ as $(.0145, .01715)$ (solid line), $(.0142, .01779)$ (dotted line) and $(.014, .01882)$ (dotdashed line).

**Case 2:** $\Gamma = \Gamma_0 \frac{T^m}{\phi^{m-1}}$

The scalar field can be written as

$$\phi^{3-m} = \left[\frac{4(3-m)}{8n - 4 - m(n-2)}\right]^2 \frac{2\delta^2 n^2 (1-n)}{3\Gamma_0 \left(\frac{\delta n(1-n)}{2}\right)^{\frac{m}{4}}} t^{\frac{8n-4-m(n-2)}{4}} = \phi_0 t^{\frac{8n-4-m(n-2)}{4}} \qquad (64)$$

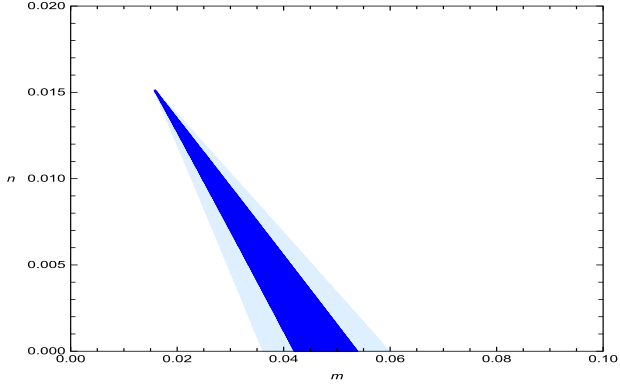

Fig. 6. In the strong dissipative regime of the present fractal warm inflationary model the observed region in the figure indicates the location of the $(n, m)$ points by using the numerical values of $(r, n_s)$ parameters from Planck 2018 data set.

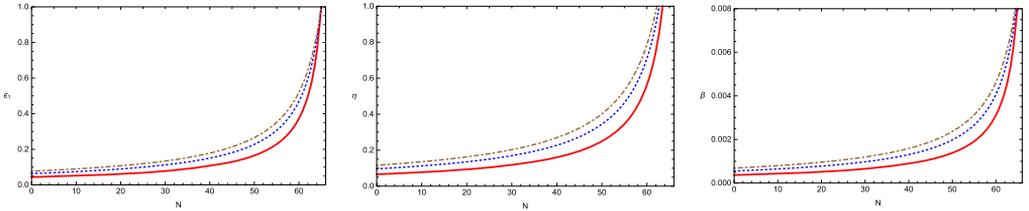

Fig. 7. Variation of slow roll parameters with respect to number of e-folds.

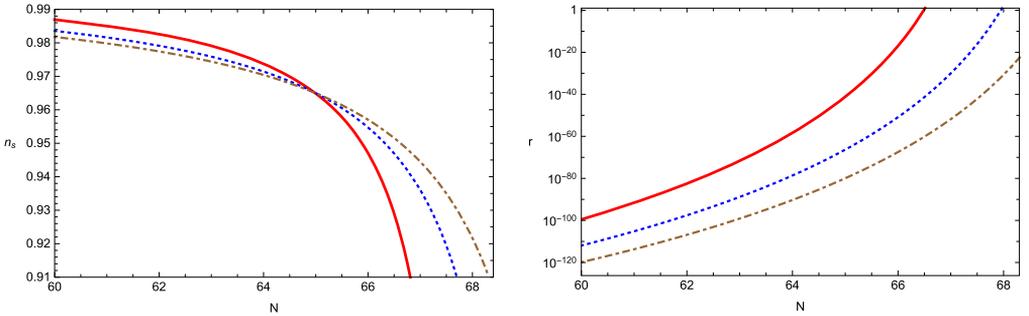

Fig. 8. Spectral index and tensor to scalar ratio versus number of e-folds.

The temperature can be written as

$$T = \left[\frac{\delta n(1-n)}{2}\right]^{\frac{1}{4}} \left[\frac{\phi^{3-m}}{\phi_0}\right]^{\frac{n-2}{8n-4-m(n-2)}} \tag{65}$$

So the dissipation coefficient $\Gamma$ can be rewritten as

$$\Gamma = \Gamma_0 \left[\frac{\delta n(1-n)}{2}\right]^{\frac{m}{4}} \frac{\phi^{\frac{m(3-m)(n-2)}{8n-4-m(n-2)}-(m-1)}}{\phi_0^{\frac{m(n-2)}{8n-4-m(n-2)}}} \tag{66}$$

Using the above relation, one can write the potential, Hubble parameter, scale factor as

$$V(\phi) = \frac{\delta^2 n^2}{3} \left[ \frac{\phi^{3-m}}{\phi_0} \right]^{\frac{8(n-1)}{8n-4-m(n-2)}} \tag{67}$$

$$H(\phi) = \frac{\delta n}{3} \left[ \frac{\phi^{3-m}}{\phi_0} \right]^{\frac{4(n-1)}{8n-4-m(n-2)}} \tag{68}$$

$$a(\phi) = a_0 \exp\left( \frac{\delta}{3} \left[ \frac{\phi^{3-m}}{\phi_0} \right]^{\frac{4n}{8n-4-m(n-2)}} \right) \tag{69}$$

The slow roll parameters can be written as

$$\epsilon_1 = \frac{3(1-n)}{\delta n} \frac{\phi^{\frac{4n(3-m)}{m(n-2)+4-8n}}}{\phi_0^{\frac{4n}{m(n-2)+4-8n}}} \tag{70}$$

$$\eta = \frac{3mn - 2m - 4n + 8}{4(m-3)(1-n)} \epsilon_1 \tag{71}$$

$$\beta = \frac{8n - 6mn - 4}{4(3-m)(1-n)} \epsilon_1 \tag{72}$$

At the end of inflation $\epsilon_1(\phi_e) = 1$, hence $\phi_e^{\frac{4n(3-m)}{m(n-2)+4-8n}} = \frac{\delta n}{3(1-n)} \phi_0^{\frac{4n}{m(n-2)+4-8n}}$. The scalar field at horizon crossing time can be written as

$$\phi_\star^{\frac{4n(3-m)}{8n-4-m(n-2)}} = \frac{3(1-n)}{n\delta} \phi_0^{\frac{4n}{8n-4-m(n-2)}} \left( 1 + \frac{n}{n-1} N \right).$$

The slow roll parameter at the horizon crossing time can be written as

$$\epsilon_{1\star} = \left( 1 + \frac{n}{n-1} N \right)^{-1} \tag{73}$$

$$\eta_\star = \frac{3mn - 2m - 4n + 8}{4(m-3)(1-n)} \left( 1 + \frac{n}{n-1} N \right)^{-1} \tag{74}$$

$$\beta_\star = \frac{8n - 6mn - 4}{4(3-m)(1-n)} \left( 1 + \frac{n}{n-1} N \right)^{-1} \tag{75}$$

The scalar spectral index can be obtained as

$$n_{s_\star}(n, m, N) = 1 - \frac{3(m(27n+8) - 26n - 2)}{8(m-3)(1-n)} \left( 1 + \frac{n}{n-1} N \right)^{-1} \tag{76}$$

Fig. 9 indicates the location of $(n, m)$ points in the observational region for the strong dissipative regime case of the present fractal warm inflationary model, considering the numerical values of $(r, n_s)$ parameters from Planck 2018 data set.

One can see that the equation (66) can also be rewritten as

$$\Gamma = \tilde{\Gamma}_0 \phi^{\tilde{m}}$$

which implies that also in SDR, case 2 can be simplified to case 1 in SDR with some modified values of $n$ and $m$. So further detailed study of graphical representation has not been done in this section.

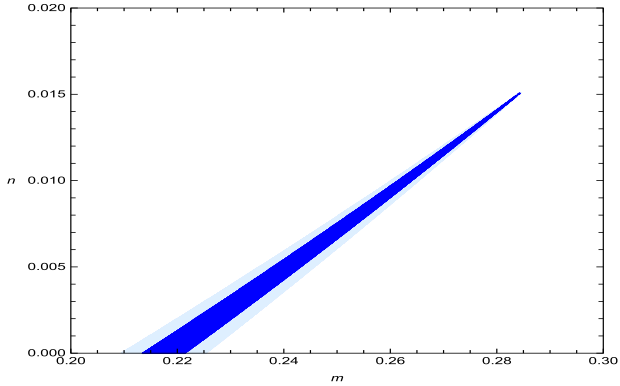

Fig. 9. The shaded region in the figure stands for the location of the $(n, m)$ point of the present model in the strong dissipative regime using the numerical values of $(r, n_s)$ from Planck-2018 data set.

## 5. Summary and conclusion

In the present work, warm inflationary scenario in the fractal gravity theory has been studied. For warm inflation, the universe is assumed to be consisting of a scalar field (inflaton) and radiation which are interacting with each other. In the fractal gravity theory, the modified Friedmann equations can be considered as Friedmann equations of Einstein gravity with interacting two fluids in which one is usual fractal fluid and other is effective fluid. Without loss of generality, in the context of warm inflation, the fractal fluid is chosen as radiation and the effective fluid has been considered as the inflaton field. Now incorporating the quasi-stable conditions and slow roll approximations the Hubble parameter, potential function and the scale factor has been expressed as a function of fractal function. Now different choices of fractal function has been considered so that it correctly describe the inflationary era. It is shown that for power law choice of fractal function inflation continues forever while exponential model of fractal function represents de-Sitter model. From the analysis it is found that both the usual choices of the fractal function are not favourable for warm inflation. So a generalized form of exponential fractal function is chosen so that it describes the inflationary paradigm and makes a smooth transition to radiation era. It is to be noted that in ref. [48] inflationary scenario with multifractal space-time geometry has been studied for various choices of the fractal parameter as a function of no. of e-folds.

For this choice of fractal function inflationary scenario have been discussed in both weak and strong dissipative regime. It is shown that in WDR, the scalar field, potential, Hubble parameters, scale factor, no. of e-fold and the first two slow roll parameter can be expressed explicitly without choosing the dissipation coefficient. To obtain expression for temperature and the third slow roll parameter one need to choose the form of dissipation coefficient. At first, the dissipation coefficient has been chosen as a function of the potential alone. Hence the scalar spectral index and tensor to scalar ratio also can be expressed as function of model parameters $(n, m)$. Using the $r - n_s$ diagram of Planck 2018 data, the allowable range for $(n, m)$ has been find out. The variations of slow roll parameters with respect to the no. of e-folds has been shown graphically which match with the slow roll approximations. Next the dissipation coefficient is chosen as a function of both temperature and scalar field. Though it is function of both temperature and scalar field, it can be shown that using the temperature expression, the dissipation coefficient can be reduced to function of potential alone. As a consequence, a similar study as previously discussed can be made with that choice for dissipation coefficient. Now for the SDR, as dissipation coefficient is dominant over Hubble parameter, one have to choose the form of dissipation coefficient at very first step.

From the above analysis, it can be concluded that whatever be the choice of the dissipation coefficient, using the quasi-stable condition it can be reduced to a function of scalar field alone. So in this method, the theoretical models can be compared with observational data by constraining the arbitrary parameters. Though the initial conditions and the birth of our universe is not completely known, however, this technique may open some possibilities to compare the predictions of the theoretical models of inflation in different modified gravity theories with cosmological observational data.

**CRediT authorship contribution statement**

**A. Bose:** Methodology, Computing, Writing, Programming, Original Data Preparation. **S. Chakraborty:** Conceptualization, Reviewing, Editing, Supervision.

**Declaration of competing interest**

The authors declare that they have no known competing financial interests or personal relationships that could have appeared to influence the work reported in this paper.

**Acknowledgement**

The author A.B. acknowledges UGC-JRF (ID: 1207/CSIRNETJUNE2019) and S.C. thanks Science and Engineering Research Board (SERB), India for awarding MATRICS Research Grant support (File No. MTR/2017/000407).